\def\review{0} 
\def\arxivdisclaimer{1} 

\documentclass[10pt, conference, letterpaper]{IEEEtran}
\usepackage[noadjust]{cite}
\usepackage{amsmath,amssymb,amsfonts}
\usepackage{graphicx}
\usepackage{textcomp}
\usepackage{mathtools}
\usepackage{adjustbox}
\usepackage{balance}
\usepackage{bm}
\usepackage{tikz}
\usetikzlibrary{arrows, fit, matrix, positioning, shapes, backgrounds}
\usepackage{xcolor}
\def\BibTeX{{\rm B\kern-.05em{\sc i\kern-.025em b}\kern-.08em
    T\kern-.1667em\lower.7ex\hbox{E}\kern-.125emX}}

\usepackage{textcase}
\usepackage[tablename=TABLE,font=small]{caption}
\DeclareCaptionTextFormat{up}{\MakeTextUppercase{#1}}
\usepackage{paralist}
\usepackage{supertabular}    
\usepackage{array} 
\usepackage[linesnumbered,ruled]{algorithm2e}
\usepackage{algorithmic}

\usepackage[acronym]{glossaries}

\if\review1
\usepackage{todonotes}
\else
\usepackage[disable]{todonotes}
\fi

\if\arxivdisclaimer0
\else
\usepackage{hyperref}
\glsdisablehyper
\fi
\usepackage{cleveref}

\usepackage{numprint}     
\usepackage{tikz}
\usepackage{makecell}   
\usepackage{xcolor,colortbl}
\usepackage{enumitem}            
\usepackage{longtable} 
\usepackage{wrapfig}
\usepackage{pgfplots}
\usepgfplotslibrary{groupplots}
\usepackage{booktabs}
\usepackage{graphics}
\usepackage{makecell}
\usetikzlibrary{spy,backgrounds}

\usepackage{tikz}
\usepackage{pgfplots}
\usepackage{pgfplotstable}
\usepackage{caption}
\usepackage{subcaption}

\pgfplotsset{compat=1.18} 

\usepackage{amsthm}
\usepackage{amsmath, amssymb}
\usepackage{tabu}

\DeclareMathOperator*{\argmax}{arg\,max}

\DeclarePairedDelimiter\ceil{\lceil}{\rceil}

\newtheorem*{remark*}{Remark}

\usepackage[utf8]{inputenc}

\crefname{figure}{Fig.}{Fig.}
\crefname{table}{Table}{Table}

\let\oldtabular\tabular
\renewcommand{\tabular}{\small\oldtabular}

\newcommand{\egc}{e.\,g., }
\newcommand{\iec}{i.\,e., }

\newcolumntype{?}{!{\vrule width 1pt}}
\definecolor{mittelblau}{RGB}{0, 126, 198}
\definecolor{violettblau}{cmyk}{0.9, 0.6, 0, 0}
\definecolor{rot}{RGB}{238, 28 35}
\definecolor{apfelgruen}{RGB}{140, 198, 62}
\definecolor{gelb}{RGB}{1, 221, 0}
\definecolor{orange}{RGB}{244, 111, 33}
\definecolor{pink}{RGB}{237, 0, 140}
\definecolor{lila}{RGB}{128, 10, 145}
\definecolor{hellgrau}{RGB}{224, 224, 224}
\definecolor{mittelgrau}{RGB}{128, 128, 128}
\definecolor{dunkelgrau}{RGB}{80,80,80}
\definecolor{anthrazit}{RGB}{19, 31, 31}
\definecolor{darkgreen}{RGB}{0.125,0.5,0.169}

\newcommand\blfootnote[1]{%
  \begingroup
  \renewcommand\thefootnote{}\footnote{#1}%
  \addtocounter{footnote}{-1}%
  \endgroup
}

\begin{document}

\title{CRAP: Clutter Removal with Acquisitions Under Phase Noise}

\author{
    \IEEEauthorblockN{
        Marcus Henninger\IEEEauthorrefmark{1}\IEEEauthorrefmark{2},
        Silvio Mandelli\IEEEauthorrefmark{1},
        Artjom Grudnitsky\IEEEauthorrefmark{1}, 
        Thorsten Wild\IEEEauthorrefmark{1},
        Stephan ten Brink\IEEEauthorrefmark{2}
        }

	\IEEEauthorblockA{
	\IEEEauthorrefmark{1}Nokia Bell Labs Stuttgart, 70469 Stuttgart, Germany \\
	\IEEEauthorrefmark{2}Institute of Telecommunications, University of Stuttgart, 70569 Stuttgart, Germany \\
	E-mail: marcus.henninger@nokia.com}}

\maketitle

\newacronym{5G}{5G}{fifth generation}
\newacronym{6G}{6G}{sixth generation}
\newacronym{agv}{AGV}{automated guided vehicle}
\newacronym{awgn}{AWGN}{additive white Gaussian noise}
\newacronym{cfar}{CFAR}{constant false alarm rate}
\newacronym{crap}{CRAP}{Clutter Removal with Acquisitions Under Phase Noise}
\newacronym{csi}{CSI}{channel state information}
\newacronym{dl}{DL}{downlink}
\newacronym{eca}{ECA}{Extensive Cancellation Algorithm}
\newacronym{eca-c}{ECA-C}{Extensive Cancellation Algorithm by Subcarrier}
\newacronym{eca-s}{ECA-S}{Extensive Cancellation Algorithm by Symbol}
\newacronym{dft}{DFT}{Discrete Fourier Transform}
\newacronym{idft}{IDFT}{Inverse Discrete Fourier Transform}
\newacronym{gnb}{gNB}{gNodeB}
\newacronym{isac}{ISAC}{Integrated Sensing and Communication}
\newacronym{kf}{KF}{Kalman Filter}
\newacronym{mdl}{MDL}{Minimum Description Length}
\newacronym{mmw}{mmWave}{millimeter-wave}
\newacronym{ofdm}{OFDM}{orthogonal frequency-division multiplexing}
\newacronym{poc}{PoC}{proof of concept}
\newacronym{ptp}{PTP}{Precision Timing Protocol}
\newacronym{rmse}{RMSE}{root-mean-square error}
\newacronym{rf}{RF}{Radio frequency}
\newacronym{ru}{RU}{Radio Unit}
\newacronym{rx}{RX}{receiver}
\newacronym{svd}{SVD}{singular value decomposition}
\newacronym{snr}{SNR}{signal-to-noise ratio}
\newacronym{synce}{SyncE}{Synchronous Ethernet}
\newacronym{tdd}{TDD}{Time Division Duplex}
\newacronym{tx}{TX}{transmitter}
\newacronym{ul}{UL}{uplink}

\begin{abstract}
The emergence of \gls{isac} in future 6G networks comes with a variety of challenges to be solved. One of those is clutter removal, which should be applied to remove the influence of unwanted components, scattered by the environment, in the acquired sensing signal. 
While legacy radar systems already implement different clutter removal algorithms, \gls{isac} requires techniques that are tailored to the envisioned use cases and the specific challenges that communications deployments bring along, like phase noise due to clock errors between transmitter and receiver.
To that end, in this work we introduce \gls{crap}. 
We propose to vectorize the time-frequency channel acquired in a radio frame in a high-dimensional space. In an offline clutter acquisition step, \acrlong{svd} is used to determine the major clutter components. At runtime, the clutter is then estimated and removed by a subspace projection of the acquired radio frame onto the clutter components.

Simulation results prove that \gls{crap} offers benefits over prior art techniques robust to phase noise. In particular, our proposal does not suppress zero Doppler information, thereby enabling the detection of slow targets. Moreover, we show \gls{crap}'s real-time applicability in a millimeter-wave \gls{isac} \acrlong{poc}, where a pedestrian is tracked in a cluttered lab environment.
\end{abstract}

\if\arxivdisclaimer1
\blfootnote{This work has been submitted to the IEEE for possible publication. Copyright may be transferred without notice, after which this version may no longer be accessible.}
\else
\vspace{0.25cm}
\fi

\glsresetall


\section{Introduction}\label{sec:intro}
\gls{isac} will be one of the main new features of future \gls{6G} systems~\cite{wild2021joint} and is already part of discussions about later \gls{5G} releases. Current research is focusing, among others, on fundamental limits, use case definition, candidate waveforms, and spectrum usage~\cite{mandelli2023survey, zhang2021enabling}. 
Whereas from an algorithmic perspective many approaches can be derived from conventional radar techniques, the peculiarities of communication systems together with the emerging use cases pose unique challenges and open up interesting research directions~\cite{tan2021integrated}.  

One such algorithmic task is clutter removal, which is applied to mitigate the effects of unwanted reflections due to objects in the environment that are not of interest for the sensing task. As this represents a long-known problem within the radar community, it has already been well researched in the past. 
Hence, various techniques exist, \egc based on Moving Target Indication~\cite{ash2018application} or Space-Time Adaptive Processing~\cite{sen2012ofdm}, to just name a few. Moreover, the \gls{eca} has been proposed~\cite{colone2009multistage}, which works by projecting the signal into a subspace orthogonal to the clutter. 
In \cite{zhao2012multipath}, \gls{eca-c} has been introduced, which applies \gls{eca} to each carrier of a multi-carrier system, and can therefore in principle also be readily deployed in \gls{ofdm} radar systems. 

However, most legacy radar operations differ significantly from the envisioned \gls{isac} use cases~\cite{zhao2012multipath}. For instance, conventional radars are usually only interested in (fast) moving objects, such as planes in airborne radar. Prior art clutter removal techniques therefore typically accept -- or even intentionally create -- suppression at zero Doppler~\cite{liu2019evaluation}, facilitating the detection of moving targets. Other approaches assume certain characteristics of the clutter, \egc sea waves in \cite{islam2012artificial}, leveraging artificial intelligence to identify clutter.
%
%
Those properties are clearly undesirable for many sensing applications currently being discussed in the context of \gls{isac}, \egc for tracking objects that are moving slowly or may even remain static for periods (humans, \glspl{agv}), with clutter of unknown characteristics.
\\In addition, classical communication setups have conceptual differences to conventional radars. For instance, the passive radars considered in \cite{colone2009multistage} and~\cite{zhao2012multipath} deploy a reference antenna steered towards the transmitter, while a surveillance antenna is steered into the direction to be surveyed by the system. In communication systems, such a reference is typically not available, rendering it difficult to adaptively estimate the clutter components. As a workaround, we rely on recording a reference scenario in an acquisition step, where only clutter components are present. To reduce the impact of noise and to capture multiple clutter components, several snapshots of the reference scenario are required. This, however, introduces phase incoherence, since in common communications hardware \gls{tx} and \gls{rx} are not sharing the same local oscillator, posing another challenge that prior art clutter removal approaches have typically not been faced with. 


This work addresses the aforementioned shortcomings with \gls{crap}. The main contribution are summarized as follows:

\begin{itemize}
    \item We propose to perform clutter acquisition and removal by acquiring multiple \gls{ofdm} symbols in a unique ``radio frame''. This is compliant with the current \gls{5G} - and likely \gls{6G} - frame structure. The radio frame is used to estimate the frame's \gls{csi} and vectorized, generating a multi-dimensional space. Multiple clutter acquisitions are then stacked in a matrix and used to determine the strongest clutter components using the \gls{svd}.
    \item The clutter is removed at runtime by subtracting, from the acquired radio frame, the projection of the acquired frame onto the clutter components.
    \item Exploiting the associative property of matrix multiplication, the computational complexity at runtime is drastically reduced, making the algorithm feasible for practical implementation.

    \item We demonstrate the benefits of \gls{crap} both with simulation results, as well as with real-world measurements. Leveraging the \gls{poc} shown in~\cite{wild2023integrated}, pedestrian tracking results in a lab environment are shown, where \gls{crap} has been applied in conjunction with a \gls{kf}.
\end{itemize}

The rest of the paper is structured as follows: in Section~\ref{sec:sys_model}, we introduce the system model, before explaining \gls{crap} in detail in Section~\ref{sec:crap}. Section~\ref{sec:sim_results} demonstrates the benefits of \gls{crap} by means of simulations. After presenting pedestrian tracking results obtained with our \gls{isac} \gls{poc} in Section~\ref{sec:poc}, the paper is concluded in Section~\ref{sec:conclusion} with a summary and a brief outlook at future work.
\section{System Model}\label{sec:sys_model}


To better introduce the problem at hand and to lay the necessary theoretical foundation, we first define a generic system model that can later also be linked to the \gls{isac} \gls{poc} in Section~\ref{sec:poc}. 
\\Consider a setup with separate \gls{tx} and \gls{rx}, where the \gls{tx} transmits $M$ \gls{ofdm} symbols at carrier frequency $f_c$, with $N$ subcarriers spaced by $\Delta f$ carrying complex modulation symbols. Accordingly, the transmitted \gls{ofdm} frame is represented by the matrix $\mathbf{X} \in \mathbb{C}^{N \times M}$. For simplicity, we consider co-located \gls{tx} and \gls{rx} in the following. Note, however, that our approach does not rely on a specific model, but can also be applied to bi-static or even distributed scenarios.
Each object in the environment indexed $p~\in~\mathcal{P}$ is characterized by its range $r_p$ and velocity  $v_p$ relative to the \gls{rx}. These objects are partitioned into the sets $\mathcal{P}_c \subset \mathcal{P}$ and $\mathcal{P}_t \subset \mathcal{P}$, comprising clutter components and targets, respectively.
The received frame $\mathbf{Y} \in \mathbb{C}^{N \times M}$ at the \gls{rx} can then be seen as the superposition of all reflections of the signal with the objects in the environment. Assuming knowledge of the transmitted symbols $\mathbf{X}$ at the \gls{rx}, the time-frequency \gls{csi} matrix is obtained via element-wise division of $\mathbf{Y}$ by $\mathbf{X}$ and given as
\begin{align}
\mathbf{H} = \phi \sum_{p \in \mathcal{P}} \alpha_{p}\mathbf{a}(r_p)\mathbf{b}(v_p)^\text{T} + \textbf{Z},
\label{eq:Y}	
\end{align} 
with $\mathit{\alpha_{p}}$ being the complex coefficient of the $\mathit{p}$-th reflection. Further, $\mathbf{Z} \in \mathbb{C}^{N \times M}$ denotes the random complex \gls{awgn} matrix, where we define each element to have a power of $\sigma_n^2 = P_n/N$, with $P_n$ representing the noise power over the full bandwidth. The vectors $\mathbf{a}(\mathit{r_{q}})$ and $\mathbf{b}(v_p)$ are written as
\begin{align}
\mathbf{a}(r_{p}) &= \begin{bmatrix}
       1, \ e^{-j4\pi \Delta f \cdot r_p/c}, \ \dots, \ e^{-j4\pi (N - 1) \Delta f \cdot r_p/c}
\end{bmatrix}^\text{T}  \\
\mathbf{b}(v_p) &= \begin{bmatrix}
       1, \ e^{j4\pi T_0 f_c \cdot v_p / c}, \ \dots, \ e^{j4\pi (M - 1) T_0  f_c \cdot v_p / c}
\end{bmatrix}^\text{T} 
\label{eq:channel_vectors}
\end{align}
where $c$ is the speed of light and $T_0$ the \gls{ofdm} symbol duration. Moreover, $\phi$ denotes the random phase rotation, which is modeled to be constant within a single sensing acquisition $\mathbf{H}$. This can be safely assumed if the phase noise random walk's standard deviation during an acquisition of one frame is $\sigma_\phi < 0.1$ rad~\cite{mandelli2014modeling}, which is typically the case in the \gls{isac} demonstrator based on communications hardware we introduced in~\cite{wild2023integrated}. Note that the phase rotation between consecutive frames does not need to be constrained to any value, nor model, in order for our approach to be valid.


Information about the objects' velocities $v_p$ and ranges $r_p$ is contained in $\mathbf{H}$ in the form of linear phase shifts in time and frequency domain, respectively. The \gls{csi} matrix can thus in principle be readily processed using conventional \gls{ofdm} radar techniques~\cite{braun2010maximum} to extract knowledge about reflective objects in the environment. However, as previously mentioned, reflections from the unwanted components in the set $\mathcal{P}_c$, such as reflections caused by (typically static) objects which are not of interest for the sensing task, are also included in the sensing acquisition. Their influence should be removed, calling for the application of clutter removal algorithms.
\section{CRAP}\label{sec:crap}
This section describes our proposed solution to the previously outlined clutter removal problem in detail. Our algorithm, \gls{crap}, can be seen as a two-step approach. 
\\First, in an ``offline" acquisition step, the matrices required for clutter removal are generated by recording multiple snapshots of the reference scenario (\iec where only clutter is present) and then extracting the meaningful clutter components. During runtime, those matrices are utilized to compute the clutter-rejected channel building upon the principles of \gls{eca}~\cite{colone2009multistage}.

\subsection{Clutter Acquisition (Offline)}\label{subsec:offline}

To capture multiple clutter components and to reduce the impact of noise, $K$ snapshots of the reference scenario are initially recorded. After vectorizing those $K$ \gls{csi} matrices, the resulting vectors of length $Q=NM$ are stacked along the rows to form the matrix $\mathbf{C}~\in~\mathbb{C}^{K \times Q}$ comprising all vectorized clutter acquisitions. Notice that doing so merely collapses the representation of the clutter into a vector, but preserves all the information, such that later clutter projection and removal can exploit the multi-dimensional space of the \gls{csi}.
In contrast, \gls{eca-c}~\cite{zhao2012multipath} operates across a \textit{single} domain, the time domain, thereby suppressing zero Doppler information when removing the clutter, as shown in Section~\ref{sec:poc}.

To relax the computational complexity at runtime, only meaningful clutter components should be considered. To that end, the \gls{svd} of the clutter matrix $\mathbf{C}$ is performed as
\begin{align}
\mathbf{C} = \mathbf{U} \mathbf{\Sigma} \mathbf{V}^{\text{H}} \; .
\label{eq:svd}
\end{align}
The clutter order $L$ can either be chosen beforehand, or determined by means of the singular values in~$\mathbf{\Sigma}$, \egc using \gls{mdl}~\cite{rissanen1978modeling}. The right singular vectors (\iec rows) of $\mathbf{V}^{\text{H}}$ corresponding to the $L$ strongest singular values are then stacked along the columns to obtain the matrix $\hat{\mathbf{C}}~\in~\mathbb{C}^{Q \times L}$, which we consider to be the clutter subspace containing the $L$ strongest clutter components.
It is hereby crucial to remark that phase fluctuations between different clutter acquisitions do not affect the singular vectors, which allows obtaining robust representations of the clutter subspace even in case of clock incoherence between TX and RX.
\\The clutter subspace $\hat{\mathbf{C}}$ alone in principle already allows to obtain the clutter projection at runtime as 
\begin{align}
\hat{\mathbf{c}} &= \overbrace{\hat{\mathbf{C}}\bigl(\hat{\mathbf{C}}^{\text{H}}\hat{\mathbf{C}}\bigr)^{-1}\hat{\mathbf{C}}^{\text{H}}}^\text{$\mathbf{P}$} \mathbf{h}
\end{align}
where the matrix $\mathbf{P}$ projecting the vectorized sensing acquisition $\mathbf{h}$ into the clutter subspace can be computed fully offline. However, determining $\mathbf{P}$ results in a $Q \times Q$ matrix, which is typically too large to perform clutter removal in real time. We therefore leverage the associative property of matrix multiplication to avoid the explicit computation of $\mathbf{P}$ by performing the multiplications in the following order 
\begin{align}
\hat{\mathbf{c}} = \overbrace{\bigl[\hat{\mathbf{C}}\bigl(\hat{\mathbf{C}}^{\text{H}}\hat{\mathbf{C}}\bigr)^{-1}\bigr]}^\text{$\mathbf{P}^{\prime}$}\bigl(\hat{\mathbf{C}}^{\text{H}}\mathbf{h}\bigr) \; . 
\label{eq:clutter_projection}
\end{align}
Clutter projection performed as per Eq.~\eqref{eq:clutter_projection} requires the Hermitian transpose of the clutter subspace $\hat{\mathbf{C}}^{\text{H}}~\in~\mathbb{C}^{L \times Q}$ and part of the projection matrix $\mathbf{P}^{\prime}~\in~\mathbb{C}^{Q \times L}$ $\mathbf{P}^{\prime}$. 
Those matrices can be pre-computed and stored, relaxing the computational burden, as typically $L \ll Q$. This further emphasizes the necessity of isolating the strongest clutter components and keeping $L$ small. The entire procedure of \gls{crap}'s clutter acquisition step is given in Algorithm~\ref{alg:clutter_acquisition}.


\begin{algorithm}[t]
 \caption{Clutter Acquisition (Offline)}\label{alg:clutter_acquisition}
 \begin{algorithmic}[1]
 \renewcommand{\algorithmicrequire}{\textbf{Input:}}
 \renewcommand{\algorithmicensure}{\textbf{Output:}}
 \REQUIRE $K$ clutter acquisition frames, \\
 \hspace{4.2mm} \textit{optional}: clutter order $L$\\ 
 \ENSURE Clutter removal matrices $\mathbf{P}^{\prime}$, $\hat{\mathbf{C}}^{\text{H}}$ \\
  \STATE $\mathbf{C} \gets$ vectorize and stack $K$ clutter acquisitions
  \STATE $\mathbf{\Sigma}$, $\mathbf{V}^{\text{H}} \gets$ perform \gls{svd} with Eq. \eqref{eq:svd}
  \IF{L not specified}
    \STATE $L \gets $ estimate clutter order using singular values  $\mathbf{\Sigma}$
  \ENDIF
  \STATE $\hat{\mathbf{C}} \gets$ store $L$ strongest right singular vectors from $\mathbf{V}^{\text{H}}$
  \STATE $\mathbf{P}^{\prime} \gets$ compute part of projection matrix using Eq.~\eqref{eq:clutter_projection}
 \RETURN $\mathbf{P}^{\prime}$, $\hat{\mathbf{C}}^{\text{H}}$
 \end{algorithmic}
 \end{algorithm}

\subsection{Clutter Removal (Runtime)}\label{subsec:online}

During runtime, the clutter-rejected sensing acquisition $\hat{\mathbf{h}}$ is obtained by subtracting the clutter projection from the current vectorized \gls{csi}
\begin{align}
\hat{\mathbf{h}} &= \mathbf{h} - \mathbf{P}^{\prime}\bigl(\hat{\mathbf{C}}^{\text{H}}\mathbf{h}\bigr)  \; .
\label{eq:clutter_removal}
\end{align}
Reshaping $\hat{\mathbf{h}}$ into its original shape finally yields the clutter-rejected \gls{csi} matrix $\hat{\mathbf{H}}$, which is further used for \gls{ofdm} radar processing.  Algorithm~\ref{alg:clutter_removal} summarizes the clutter removal step at runtime.

Notice that our algorithm can straightforwardly be extended to further dimensions, \egc by considering also the angular domain (multiple antennas).



\begin{algorithm}[t]
 \caption{Clutter Removal (Runtime)}\label{alg:clutter_removal}
 \begin{algorithmic}[1]
 \renewcommand{\algorithmicrequire}{\textbf{Input:}}
 \renewcommand{\algorithmicensure}{\textbf{Output:}}
 \REQUIRE  Clutter removal matrices $\mathbf{P}^{\prime}$, $\hat{\mathbf{C}}^{\text{H}}$, \\ 
 \hspace{4.2mm} current sensing acquisition $\mathbf{H}$ \\
 \ENSURE Clutter-rejected sensing acquisition $\hat{\mathbf{H}}$ \\
  \STATE $\mathbf{h} \gets$ vectorize current sensing acquisition $\mathbf{H}$ 
 \STATE $\hat{\mathbf{h}} \gets$ get clutter-rejected sensing acquisition using $\mathbf{P}^{\prime}$, $\hat{\mathbf{C}}^{\text{H}}$, and Eq.~\eqref{eq:clutter_removal}
  \STATE $\hat{\mathbf{H}} \gets $ reshape clutter-rejected sensing acquisition
 \RETURN $\hat{\mathbf{H}}$
 \end{algorithmic}
 \end{algorithm}

\subsection{A note on computational complexity}

\gls{crap}'s computational bottleneck is clearly the \gls{svd} of the $K \times Q$ matrix $\mathbf{C}$ in \eqref{eq:svd}. This can however be tolerated, as the first step of our approach is performed completely ``offline". The runtime complexity of our proposed solution then scales linearly with $QL$, hence with $Q$, as typically $L \ll Q$, which has proven to be sufficient for real-time implementation. However, one could still apply decimation techniques to undersample the \gls{csi} matrix and get $Q' \ll Q$ to further reduce complexity. In that case, interpolation techniques may be applied to $\hat{\mathbf{H}}$ after clutter removal to ``recover" the information lost by under-sampling, \egc based on the principles of~\cite{mandelli2022sampling}. This could be of particular interest in case the application of the algorithm is extended to further dimensions.
\section{Simulation Results}\label{sec:sim_results}

This section will demonstrate the benefits of \gls{crap} by means of simulation results. We first describe our simulation setup and the considered baselines, which are then compared to \gls{crap} in terms of target detection and \gls{rmse} performance. 

\subsection{Simulation Setup}

For our simulation campaign, we consider the system model described in Section~\ref{sec:sys_model} with co-located \gls{tx} and \gls{rx}, each with a single antenna. 
\\We randomly place $\lvert \mathcal{P}_c \rvert = 5 $ static clutter components in the environment, which is in line with the number of multipath clusters that can be expected in an indoor environment~\cite{czink2004number}. Note that we assume $\lvert \mathcal{P}_c \rvert$ to be known such that $L$ is chosen to be $5$. As previously mentioned and also later done in our \gls{poc} experiments in Section~\ref{sec:poc}, $L$ may in practice be estimated based on the singular values using model order estimation algorithms such as \gls{mdl}. During the clutter acquisition phase, the clutter contributions are captured with $K=100$ snapshots. 

For each experiment, a single target, whose presence is to be detected by the sensing system, is added to the scenario at a random range at runtime. The target's velocities are uniformly drawn between -1.5~m/s and 1.5~m/s, corresponding to a typical pedestrian tracking use case. Note that we confine the maximum one-way distance between all objects and the radar to 25~m. The complex coefficients $\alpha_p$ of all objects are generated according to a Rician distribution, and free-space path loss is considered to model their attenuation. 
In our simulations, we also consider noise powers below the thermal noise floor $P_n^* = -174\frac{\text{dBm}}{\text{Hz}} \cdot B \approx 91$~dBm to investigate the performance at higher \glspl{snr}, which can occur in other scenarios, \egc in case of higher transmit powers, closer targets, or larger radar cross-sections.


The \gls{rf} parameters chosen for the simulation study are closely coupled to our \gls{mmw} \gls{isac} \gls{poc}. Each of the $K$ acquisitions comprises $M = 1120$ symbols, corresponding to the number of symbols per radio frame for the numerology that was also used in our demonstrator~\cite{3gpp_38211}. A summary of the simulation parameters can be found in Table~\ref{tab:sim_params}. 


\begin{table}[t]
	\caption{Simulation Parameters. \label{tab:sim_params}}
	\centering
     \begin{tabular}{|l|r|}
       		\hline
       		Number of subcarriers $N$ & 1584 \\
            \hline
            Number of \gls{ofdm} symbols per radio frame $M$ & 1120 \\
  			\hline
  			Carrier frequency $f_c$ & 27.4 GHz \\
  			\hline
  			Subcarrier spacing  $\Delta f$ & 120 kHz \\
  			\hline
            Total bandwidth $B$ & 190 MHz \\
  			\hline
            Number of clutter components $\lvert \mathcal{P}_c \rvert$ & 5 \\
            \hline
            Number of clutter acquisitions $K$ & 100 \\
  			\hline
             Probability of false alarm $P_{FA}$ & 0.001 \\
  			\hline
    \end{tabular}
\end{table}

We compare \gls{crap} to the following baselines:

\subsubsection{\textbf{``\gls{eca-c}"}} As mentioned earlier, \gls{eca-c}~\cite{zhao2012multipath} is the closest prior art and works by projecting and removing the clutter per subcarrier. This modified version of \gls{eca-c} uses the same $K$ clutter acquisitions, but does not apply vectorization and instead performs clutter removal per subcarrier. The clutter-rejected $n$-th carrier is given as
\begin{align}
    \hat{\mathbf{h}}_{\text{c}, n}^\text{T} = \mathbf{H}(n, :) - \mathbf{H}(n, :)  \mathbf{C}_{\text{c}, n}^\text{H} \bigl( \mathbf{C}_{\text{c}, n}  \mathbf{C}_{\text{c}, n}^\text{H} \bigr)^{-1} \mathbf{C}_{\text{c}, n} 
      \label{eq:clutter_rejected_sc}
\end{align}
with $\mathbf{H}(n, :)$ being the $n$-th row, \iec subcarrier, of the current sensing acquisition. Further, $\mathbf{C}_{\text{c}, n}~\in~\mathbb{C}^{K \times M}$ denotes the $n$-th clutter matrix, which is obtained by stacking the $n$-th carrier's $K$ clutter acquisition vectors of length $M$ along the rows. The full clutter-rejected sensing acquisition after applying \gls{eca-c} is then $\hat{\mathbf{H}}_{\text{ECA-C}} = \begin{bmatrix} \hat{\mathbf{h}}_{\text{c}, 1} & \hat{\mathbf{h}}_{\text{c}, 2} & \hdots & \hat{\mathbf{h}}_{\text{c}, N} 
\end{bmatrix}^{\text{T}}$.

\subsubsection{\textbf{``\acrshort{eca-s}"}} As another baseline, we propose \gls{eca-s}. This technique relies on the same principles as \gls{eca-c}, but leverages the multicarrier nature of the \gls{ofdm} signal and performs clutter removal on each \textit{symbol}, \iec it simply operates in the frequency domain instead of the time domain as \gls{eca-c}.
Hence, the clutter-rejected $m$-th symbol is
\begin{align}
    \hat{\mathbf{h}}_{\text{s}, m} = \mathbf{H}(:, m)  - \mathbf{C}_{\text{s}, m}  \bigl(\mathbf{C}_{\text{s}, m}^\text{H} \mathbf{C}_{\text{s}, m}   \bigr)^{-1} \mathbf{C}_{\text{s}, m}^\text{H}\mathbf{H}(:, m)  
\label{eq:clutter_rejected_symbol}
\end{align}

where $\mathbf{H}(:, m)$ extracts the $m$-th column (symbol) out of $\mathbf{H}$. The clutter matrix $\mathbf{C}_{\text{s}, m}~\in~\mathbb{C}^{N \times K}$ is now constructed by stacking the $m$-th symbol's $K$ clutter acquisition vectors of length $N$ along the columns. The complete clutter-rejected sensing acquisition with \gls{eca-s} is written as $\hat{\mathbf{H}}_{\text{ECA-S}} = \begin{bmatrix} \hat{\mathbf{h}}_{\text{s}, 1} & \hat{\mathbf{h}}_{\text{s}, 2} & \hdots & \hat{\mathbf{h}}_{\text{s}, M} 
\end{bmatrix}$.
  
\subsubsection{\textbf{``No Removal"}} Finally, we consider also the case without applying any clutter removal algorithm, \iec target detection is performed directly on the unprocessed sensing acquisition $\mathbf{H}$.

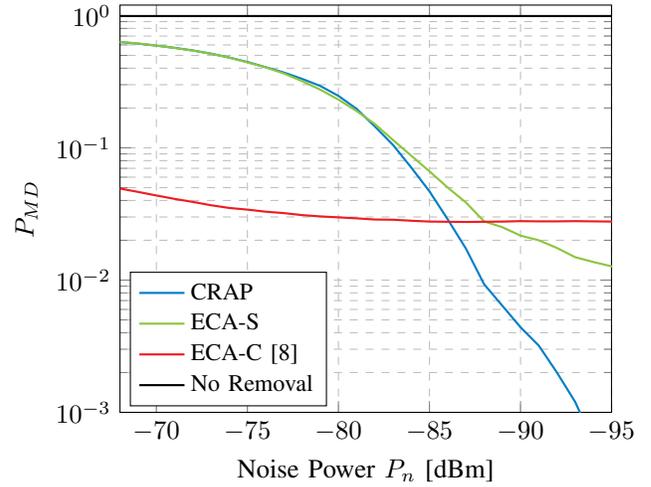
\begin{figure}[!t]
\def\scale{1}

\begin{tikzpicture}
		\begin{semilogyaxis}[
			height = 7cm,
    		xlabel={Noise Power $P_n$ [dBm]},
    		ylabel={$P_{MD}$},
            xmin = -95,
            x dir=reverse,
    		ymin=0.001,
		      ymax = 1.2,
		      enlargelimits = false,
    		xmajorgrids=true,
    		yminorgrids=true,
    		grid style=dashed,
    		legend columns = 1,
    		legend style={at={(0.02,0.18)},
            anchor=west, font=\small, 
            },
		legend cell align={left},
		every axis plot/.append style={thick},
        scale = \scale,
		]
    	  	
    	\addplot[
   		color=mittelblau,
        mark options={solid}]
    	plot table[x expr=\thisrowno{0}+30+31.998, y index=1] {Data/Missed_Det_Clutter_Components_5_Clutter_Acquisitions_100_Single.txt};
         \addlegendentry{\gls{crap}}

         \addplot[
   		color=apfelgruen,
        mark options={solid}]
    	plot table[x expr=\thisrowno{0}+30+31.998, y index=3] {Data/Missed_Det_Clutter_Components_5_Clutter_Acquisitions_100_Single.txt};
         \addlegendentry{ECA-S}

        \addplot[
   		color=rot,
        mark options={solid}]
    	plot table[x expr=\thisrowno{0}+30+31.998, y index=2] {Data/Missed_Det_Clutter_Components_5_Clutter_Acquisitions_100_Single.txt};
         \addlegendentry{ECA-C \cite{zhao2012multipath}}


         \addplot[
   		color=black,
        mark options={solid}]
    	plot table[x expr=\thisrowno{0}+30+31.998, y index=5] {Data/Missed_Det_Clutter_Components_5_Clutter_Acquisitions_100_Single.txt};
         \addlegendentry{No Removal}
    
	\end{semilogyaxis}
\end{tikzpicture} 
  \caption{Probability of missed detection for our proposal ``\gls{crap}", baselines ``\gls{eca-c}" and ``\acrshort{eca-s}", as well as ``No Removal", \iec without any clutter removal applied.} 
  \label{fig:prob_det}
\end{figure}

The respective clutter-rejected \gls{csi} matrices are then further processed using \gls{ofdm} radar principles to estimate the target's range and velocity. A maximum likelihood estimate of the parameters can be found by evaluating the periodogram, which is obtained by performing a \gls{dft} over the \gls{ofdm} symbols and an \gls{idft} over the subcarriers of $\hat{\mathbf{H}}$ \cite{braun2010maximum}
\begin{align}
S(n,m) = \frac{1}{N'M'}\bigg|\sum_{k=0}^{N'} \Biggl(\sum_{l=0}^{M'} \hat{\mathbf{H}}(k, l)e^{-j2\pi\frac{lm}{M'}}\Biggr)e^{j2\pi\frac{kn}{N'}} \bigg|^2
\label{eq:per} \;
\end{align}
where $N' = 2^{\ceil{\log_2{N}}}$ and $M' = 2^{\ceil{\log_2{M}}}$ denote the number of rows and columns of $\hat{\mathbf{H}}$ after zero padding. 
\\ As multi-peak detection and matching techniques are not the scope of this work, we restrict to the detection of only the strongest peak. Accordingly, the indices of the strongest periodogram bin are found as
\begin{align}
(\hat{n},\hat{m}) = \argmax_{n,m} S(n,m)
\label{eq:argmax} \; .
\end{align}
We consider the peak to be a detection if $S(\hat{n}, \hat{m})$ exceeds a \gls{cfar} threshold $\eta$ defined by a pre-determined probability of false alarm $P_{FA}$. The target's range $\hat{r}$ and speed $\hat{v}$ estimates are obtained by applying interpolation to the bin indices, and converting the resulting fractional indices to range and Doppler values as described in~\cite{braun2014ofdm}.

\begin{figure*}[!t]
\centering
\begin{subfigure}{0.49\textwidth}
   \def\scale{1}

\begin{tikzpicture}
		\begin{semilogyaxis}[
			height = 7cm,
    		xlabel={Noise Power $P_n$ [dBm]},
    		ylabel={Range RMSE [m]},
            x dir=reverse,
            xmin = -95,
            ytick={0.01, 0.02, 0.03, 0.04, 0.05, 0.06, 0.07, 0.08, 0.09, 0.1, 0.2, 0.3, 0.4, 0.5, 0.6, 0.7, 0.8, 0.9, 1},
            yticklabels={$10^{-2}$, , , , , , , , , $10^{-1}$, , , , , , , , ,  $10^{0}$},
    		ymin = 0.01,
		      ymax = 1,
		      enlargelimits = false,
            grid = both,
    		grid style=dashed,
    		legend columns = 2,
    		legend style={at={(0.02,0.18)},
            anchor=west, 
            font=\small},
		      legend cell align={left},
            legend columns = 1,
    		every axis plot/.append style={thick},
            scale = \scale,
    		]
    	  	
    	\addplot[
   		color=mittelblau,
        mark options={solid}]
    	plot table[x expr=\thisrowno{0}+30+31.998, y index=1] {Data/Range_RMSE_Clutter_Components_5_Clutter_Acquisitions_100_Single.txt};
         \addlegendentry{\gls{crap}}
         
        \addplot[
   		color=apfelgruen,
        mark options={solid}]
    	plot table[x expr=\thisrowno{0}+30+31.998, y index=3] {Data/Range_RMSE_Clutter_Components_5_Clutter_Acquisitions_100_Single.txt};
         \addlegendentry{ECA-S}

        \addplot[
   		color=rot,
        mark options={solid}]
    	plot table[x expr=\thisrowno{0}+30+31.998, y index=2] {Data/Range_RMSE_Clutter_Components_5_Clutter_Acquisitions_100_Single.txt};
         \addlegendentry{ECA-C \cite{zhao2012multipath}}


         \addplot[
   		color=black,
        mark options={solid}]
    	plot table[x expr=\thisrowno{0}+30+31.998, y index=5] {Data/Range_RMSE_Clutter_Components_5_Clutter_Acquisitions_100_Single.txt};
         \addlegendentry{No Removal}
    
	\end{semilogyaxis}
\end{tikzpicture} 
   \caption{Range RMSE.}
   \label{fig:range_rmse}
\end{subfigure}
\begin{subfigure}{0.49\textwidth}
   \def\scale{1}

\begin{tikzpicture}
		\begin{semilogyaxis}[
			height = 7cm,
    		xlabel={Noise Power $P_n$ [dBm]},
    		ylabel={Velocity RMSE [m/s]},
            x dir=reverse,
            xmin = -95,
            ytick={0.01, 0.02, 0.03, 0.04, 0.05, 0.06, 0.07, 0.08, 0.09, 0.1, 0.2, 0.3, 0.4, 0.5, 0.6, 0.7, 0.8, 0.9, 1},
            yticklabels={$10^{-2}$, , , , , , , , , $10^{-1}$, , , , , , , , ,  $10^{0}$},
    		ymin = 0.01,
		      ymax = 1,
		      enlargelimits = false,
            grid = both,
    		grid style=dashed,
    		legend columns = 2,
    		legend style={at={(0.02,0.18)},
            anchor=west, 
            font=\small},
		      legend cell align={left},
        legend columns = 1,
		every axis plot/.append style={thick},
        scale = \scale,
		]
    	  	
    	\addplot[
   		color=mittelblau,
        mark options={solid}]
    	plot table[x expr=\thisrowno{0}+30+31.998, y index=1] {Data/Speed_RMSE_Clutter_Components_5_Clutter_Acquisitions_100_Single.txt};
         \addlegendentry{\gls{crap}}

         \addplot[
   		color=apfelgruen,
        mark options={solid}]
    	plot table[x expr=\thisrowno{0}+30+31.998, y index=3] {Data/Speed_RMSE_Clutter_Components_5_Clutter_Acquisitions_100_Single.txt};
         \addlegendentry{ECA-S}

        \addplot[
   		color=rot,
        mark options={solid}]
    	plot table[x expr=\thisrowno{0}+30+31.998, y index=2] {Data/Speed_RMSE_Clutter_Components_5_Clutter_Acquisitions_100_Single.txt};
         \addlegendentry{ECA-C \cite{zhao2012multipath}}


         \addplot[
   		color=black,
        mark options={solid}]
    	plot table[x expr=\thisrowno{0}+30+31.998, y index=5] {Data/Speed_RMSE_Clutter_Components_5_Clutter_Acquisitions_100_Single.txt};
         \addlegendentry{No Removal}
    
	\end{semilogyaxis}
\end{tikzpicture} 
   \caption{Velocity RMSE.}
   \label{fig:speed_rmse}
\end{subfigure}

\caption{Range and velocity \gls{rmse} for our proposal ``\gls{crap}, baselines ``\gls{eca-c}" and ``\gls{eca-s}", as well as ``No Removal", \iec without any clutter removal applied.}
\label{fig:rmse}
\end{figure*}
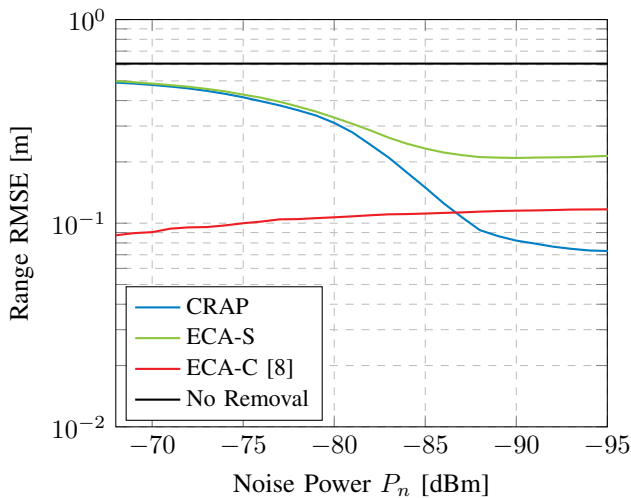
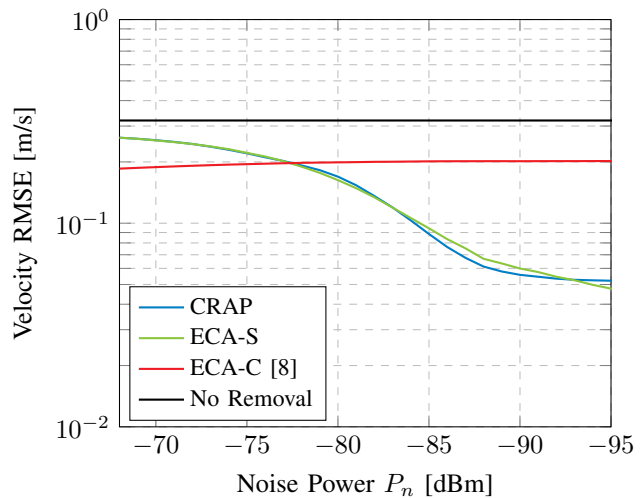

\subsection{Detection Performance}

We now evaluate \gls{crap}'s target detection capabilities compared to the previously defined baselines ``\gls{eca-c}", ``\gls{eca-s}", and ``No Removal". All approaches are evaluated in terms of the probability of missed detection $P_{MD}$, defined as the percentage of undetected targets. Note that we only consider a peak a detection if both its range and speed errors are smaller than their corresponding theoretical resolutions, which are given as $\Delta r = \frac{c}{2N\Delta f}$ and $\Delta v = \frac{c \Delta f}{2M f_c}$, respectively~\cite{braun2014ofdm}.

\cref{fig:prob_det} displays the missed detection probability curves. First, it can be noticed that without clutter removal (``No Removal") the missed detection probability is close to one, as typically a clutter component causes the strongest reflection. One can further observe that, while ``\gls{eca-c}" offers the best detection capabilities for high noise powers, its $P_{MD}$ saturates quickly. This can be attributed to the zero Doppler suppression effect, leaving (almost) static targets undetected. We will further illustrate this behavior by means of a real-world example in Section~\ref{sec:poc}. A similar saturation effect can also be noticed for ``\gls{eca-s}". These missed detections are due to ``\gls{eca-s}" operating only in the frequency domain, and occur when the target is at the same range as a strong clutter component. This can be seen as the analogue to the attenuation at zero Doppler that is created by \gls{eca-c} due to static clutter.
\\Our proposal ``\gls{crap}", on the other side, does not exhibit such a saturation effect. Once the noise power is low enough that the $K$ acquisitions allow to completely capture the clutter components, their influence can be removed without canceling the true target. Whereas the exact extents of the \gls{eca-c}/\gls{eca-s} saturation effects clearly depend on the clutter components, thus on the scenario, it can be stated that \gls{crap} is the only technique that does not impose requirements on target or clutter characteristics. Moreover, it is worth noting that the \gls{snr} in the low noise power regimes, \iec where ``\gls{crap}" performs best, can typically be assumed in sensing applications thanks to the available processing gain (due to the 2D \gls{dft}) or beamforming gains, as in real demonstrators~\cite{wild2023integrated}. 


\subsection{\acrshort{rmse} Performance}

Next, the \gls{rmse} performance is investigated. To that end, the \gls{rmse} for parameter $\Theta$ (either range $r$ or velocity $v$) is defined as 
\begin{align}
\text{RMSE} = \sqrt{E_j \Big[|\hat{\Theta}_j - \Theta_j|^2 \Big]} 
\label{eq:RMSE}
\end{align}
where $\hat{\Theta}_j$ and ${\Theta}_j$ are estimate and true value of the respective parameter in the $j$-th experiment. Note that in case of a missed detection, we assume an error equal to the respective \gls{dft} bin width (in terms of range and velocity).

\cref{fig:rmse} displays the range and velocity \gls{rmse} curves of the investigated techniques. ``\gls{crap}" converges to range and speed errors of ca. 0.07~m and 0.05~m/s, respectively, once no missed detections occur anymore.
``\gls{eca-s}" offers similar velocity estimation capabilities as ``\gls{crap}", but suffers in the range estimation. The adverse effect is observed for ``\gls{eca-c}", which offers an almost equal range estimation performance as ``\gls{crap}", but loses in velocity estimation. Again, this can be attributed to the fact that \gls{eca-c} and \gls{eca-s} perform clutter removal only in time and frequency domain, respectively, thereby causing performance degradation in the respective dimension.


\section{PoC Results}\label{sec:poc}


\begin{figure*}[t]
    \begin{center}
        \includegraphics[width=0.80\textwidth,
        trim={0mm 0mm 0mm 0mm}]{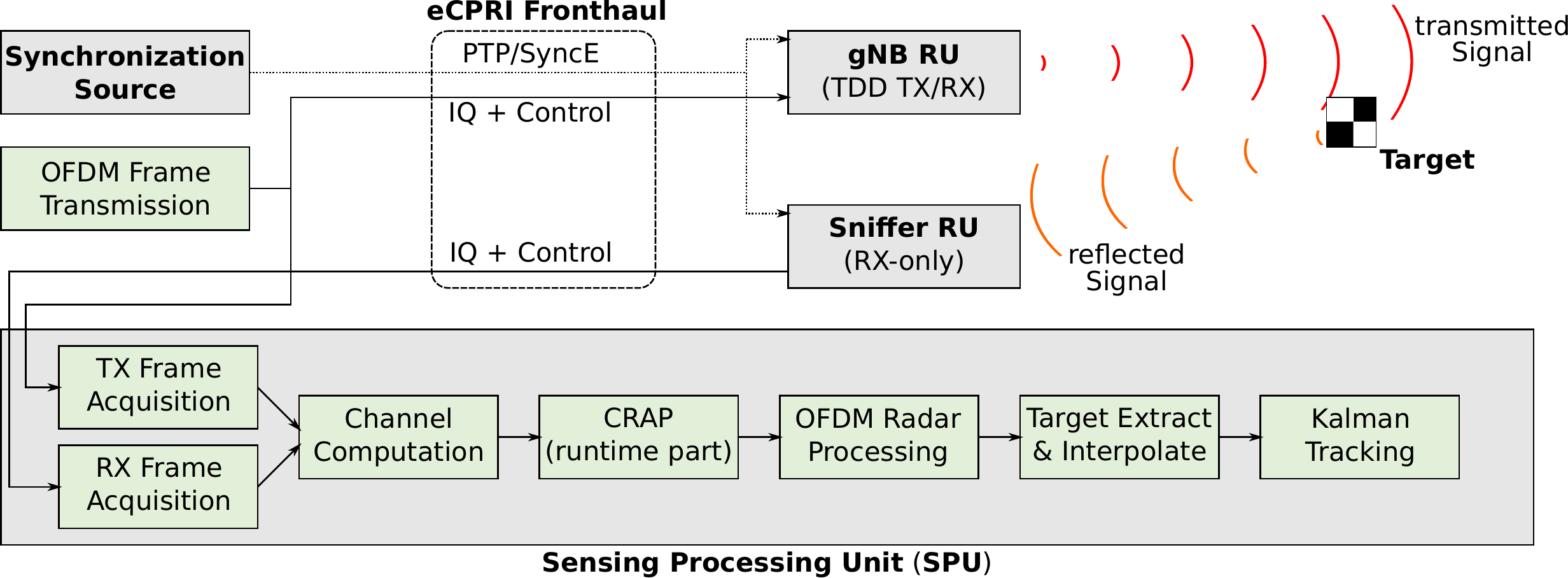}
    \end{center}
\caption{\gls{isac} \gls{poc} schematic with main processing blocks and interfaces.}
  \label{fig:sys_overview}
\end{figure*}

To evaluate \gls{crap}, we integrate it as part of our \gls{isac} \gls{poc}, and use the system to track the movement of a pedestrian.

The \gls{poc} is built on top of a 5G \gls{gnb} operating in FR2 (27.4 GHz carrier frequency) with a 200 MHz carrier and $\mu=3$ numerology, consistent with the parameters in \cref{tab:sim_params}.
We use a 64-QAM modulated downlink signal, transmitted using a 4:1 \gls{dl}/\gls{ul} frame structure~\cite{3gpp_38211}.

\begin{figure}[!b]
\centering
  \includegraphics
  [width=0.8\linewidth,
  height = 7cm,
  clip,
  trim={0mm 20mm 0mm 5mm}]
  {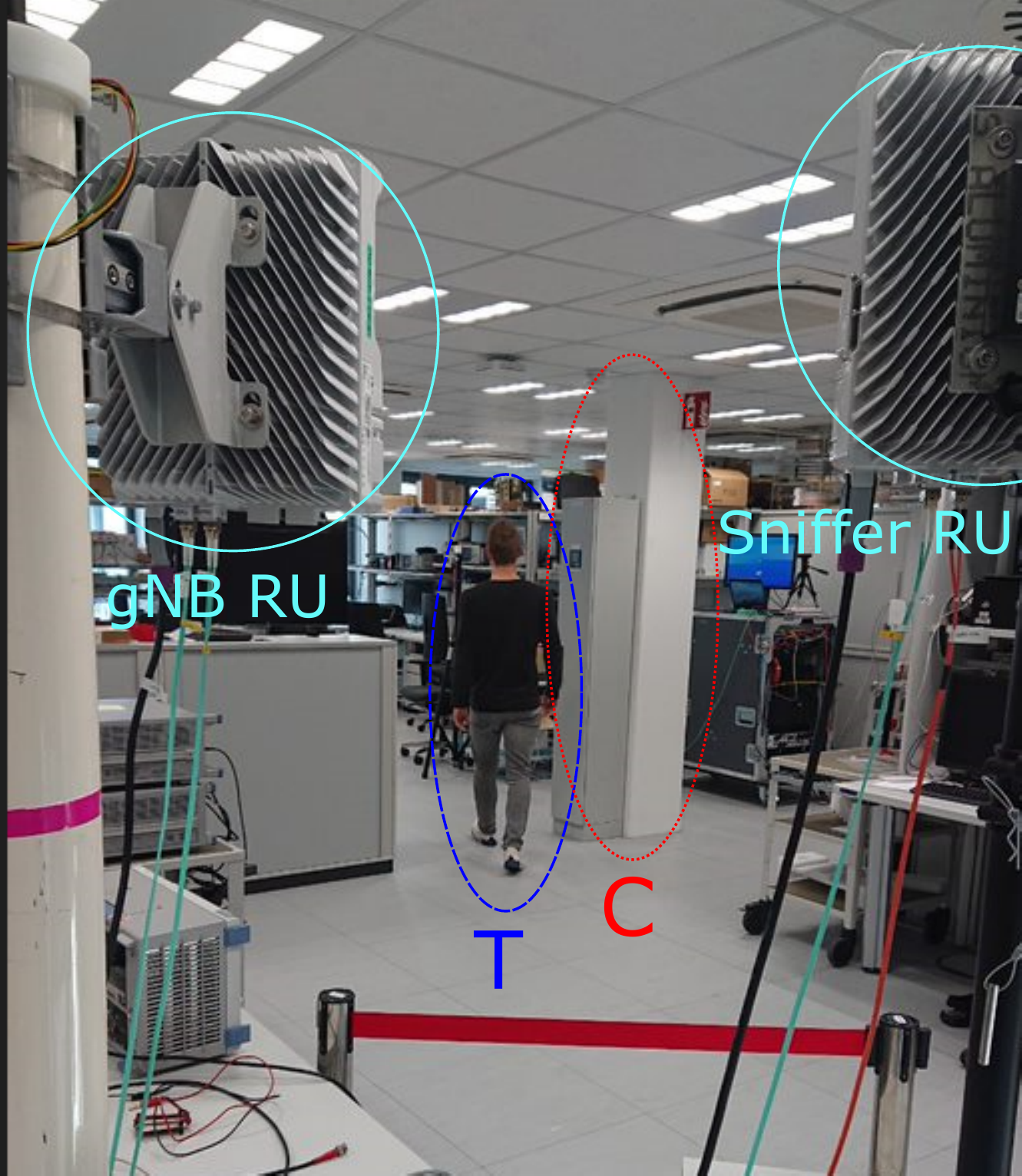}
\caption{Human target \textbf{T} walking away from the \gls{isac} \gls{poc} in the direction of the strong clutter component \textbf{C}.}
\label{fig:walking_lab}
\end{figure}

As our radio hardware is not full-duplex capable, we are using quasi co-located \gls{tx} and \gls{rx} as in \cref{fig:walking_lab}, properly isolated to avoid self-interference. Both transceivers are analog uniform rectangular arrays with $16 \times 16$ antenna elements, separated by half a wavelength.
Both \glspl{ru} are synchronized using a combination of \gls{ptp}~\cite{PTPv2_spec} and \gls{synce}~\cite{SyncE_spec}, with a measured synchronization jitter of approx. 33~ps (standard deviation).
Still, even this level of synchronization precision means that there is noticeable phase incoherence ($\phi$) between the transmitted and reflected signals, as described in the system model in \Cref{sec:sys_model}, and addressed in \Cref{subsec:offline}. In our setup, however, the phase noise standard deviation during a radio frame acquisition $\sigma_\phi$ is such that periodogram operations can be performed with negligible distortion.

We acquire the transmitted signals by monitoring the fronthaul to the \gls{gnb} \gls{ru} and stream that along with the reflected signal from the Sniffer \gls{ru} into a real-time OFDM radar, which performs the following steps:
\begin{compactitem}
\item TX/RX Radio Frame Acquisition, result: Matrices of TX/RX Signals for full radio frame
\item Channel Computation, result: Channel matrix
\item CRAP (runtime part), result: Clutter-rejected channel matrix
\item OFDM Radar Processing, result: 2D-Periodogram
\item Target Extraction and Interpolation, result: Target list
\item Kalman Tracking, result: Modified range and speed for one target
\end{compactitem}
Additionally, the implementation allows recording transmitted and reflected signals to allow (i) offline OFDM Radar Processing and (ii) Clutter Acquisition as described in \Cref{subsec:offline}. An overview of the \gls{isac} \gls{poc} with its main processing blocks is given in~\cref{fig:sys_overview}. A more detailed description of the \gls{poc} realization, can be found in~\cite{wild2023integrated}.

For tracking the pedestrian, we deploy a simple \gls{kf}~\cite{kalman1960new}, which tracks the status $\mathbf{x}_i = [{r}_i, {v}_i ]$ at time $i$, where ${r}_i$ and ${v}_i$ are the \gls{kf} posterior range and speed estimates at time $i$, respectively.
\begin{figure*}[!t]


\begin{subfigure}{0.325\linewidth}
    \begin{center}
        \includegraphics
        [width=1.0\linewidth,
        height = 5cm,
        clip,
        trim={5mm 0.5mm 5mm 14mm}]
        {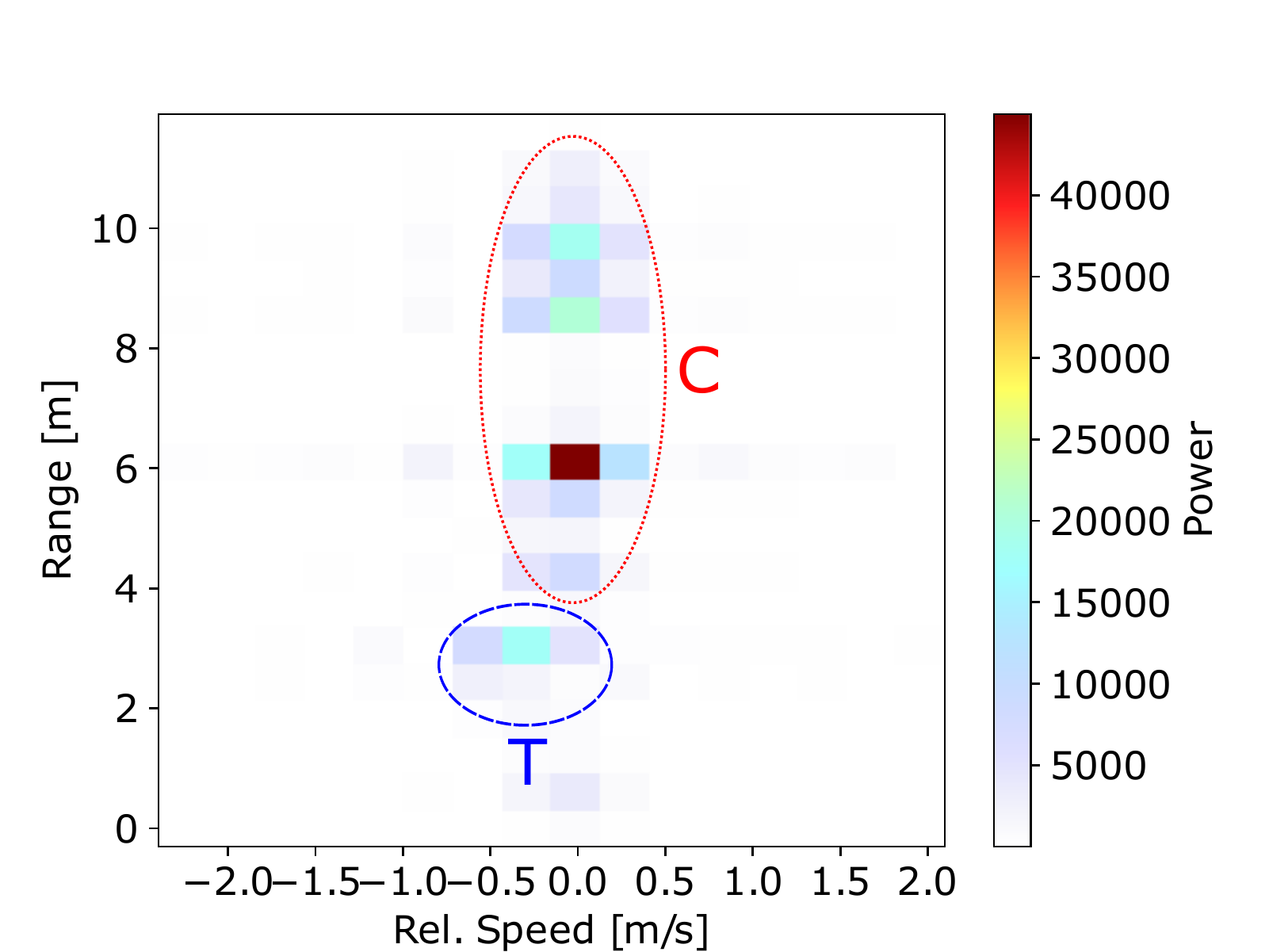} 
    \end{center}
  \caption{no clutter removal, pedestrian at range of ca.~3~m.}
  \label{fig:no_rem_3m}
\end{subfigure}
\:
\begin{subfigure}{0.325\linewidth}
    \begin{center}
        \includegraphics        [width=1.0\linewidth,
        height = 5cm,
        clip,trim={5mm 0.5mm 13mm 14mm}]
        {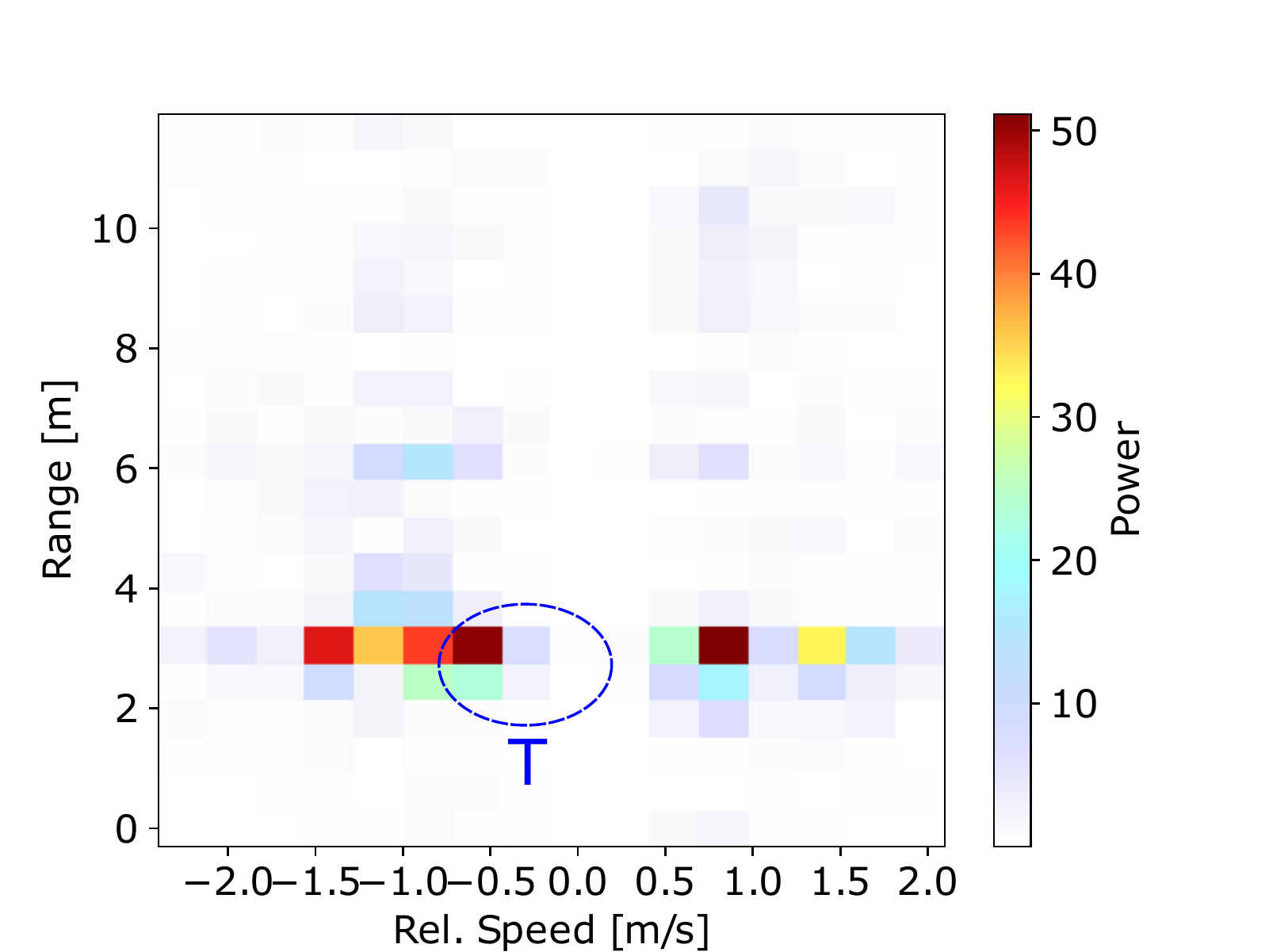} 
    \end{center}
   \caption{clutter removal with \gls{eca-c}, pedestrian at range of ca.~3~m.}
   \label{fig:eca_c_3m}
\end{subfigure}
\:
\begin{subfigure}{0.325\linewidth}
    \begin{center}
        \includegraphics        [width=1.0\linewidth,
        height = 5cm,
        clip,
        trim={5mm 0.5mm 5mm 14mm}]
        {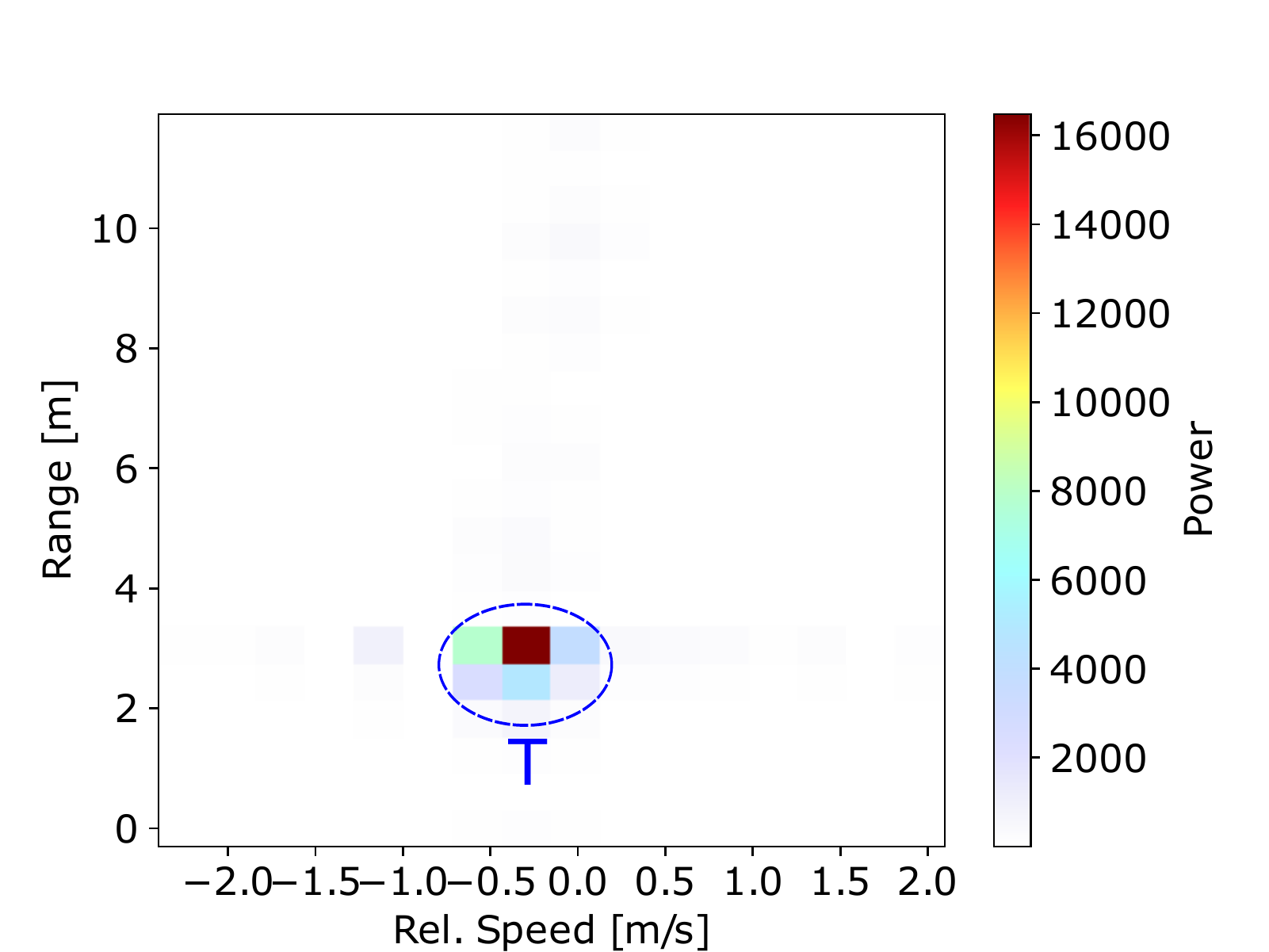} 
   \end{center}
   \caption{clutter removal with \gls{crap}, pedestrian at range of ca.~3~m.}
   \label{fig:crap_3m}
\end{subfigure}
\vskip\baselineskip


\begin{subfigure}{0.325\linewidth}
    \begin{center}
        \includegraphics
        [width=1.0\linewidth,
        height = 5cm,
        clip,
        trim={5mm 0.5mm 5mm 14mm}]
        {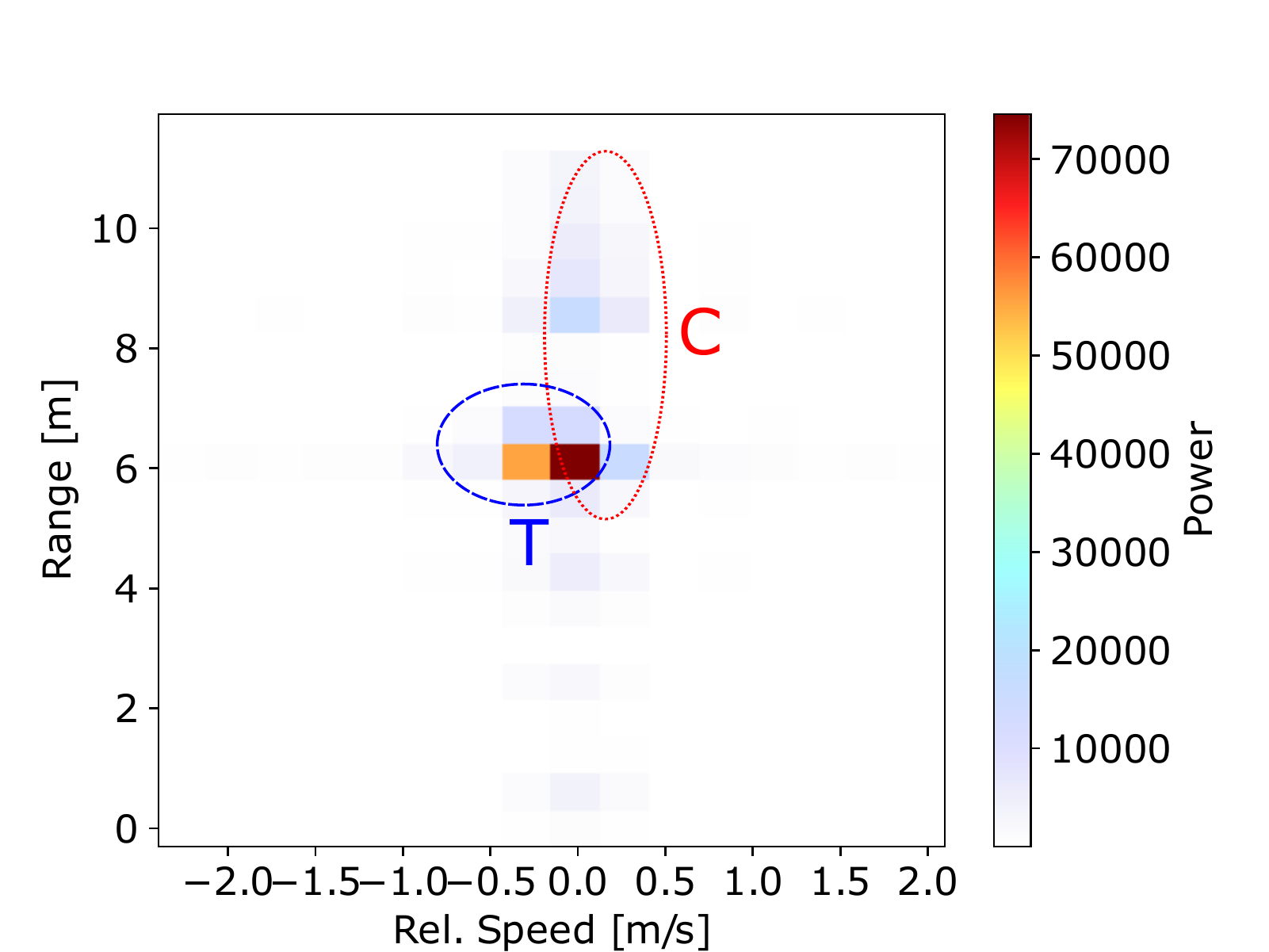} 
    \end{center}
  \caption{no clutter removal, pedestrian close to clutter at range of ca. 6~m.}
  \label{fig:no_rem_close}
\end{subfigure}
\:
\begin{subfigure}{0.325\linewidth}
    \begin{center}
        \includegraphics        [width=1.0\linewidth,
        height = 5cm,
        clip,trim={5mm 0.5mm 13mm 14mm}]
        {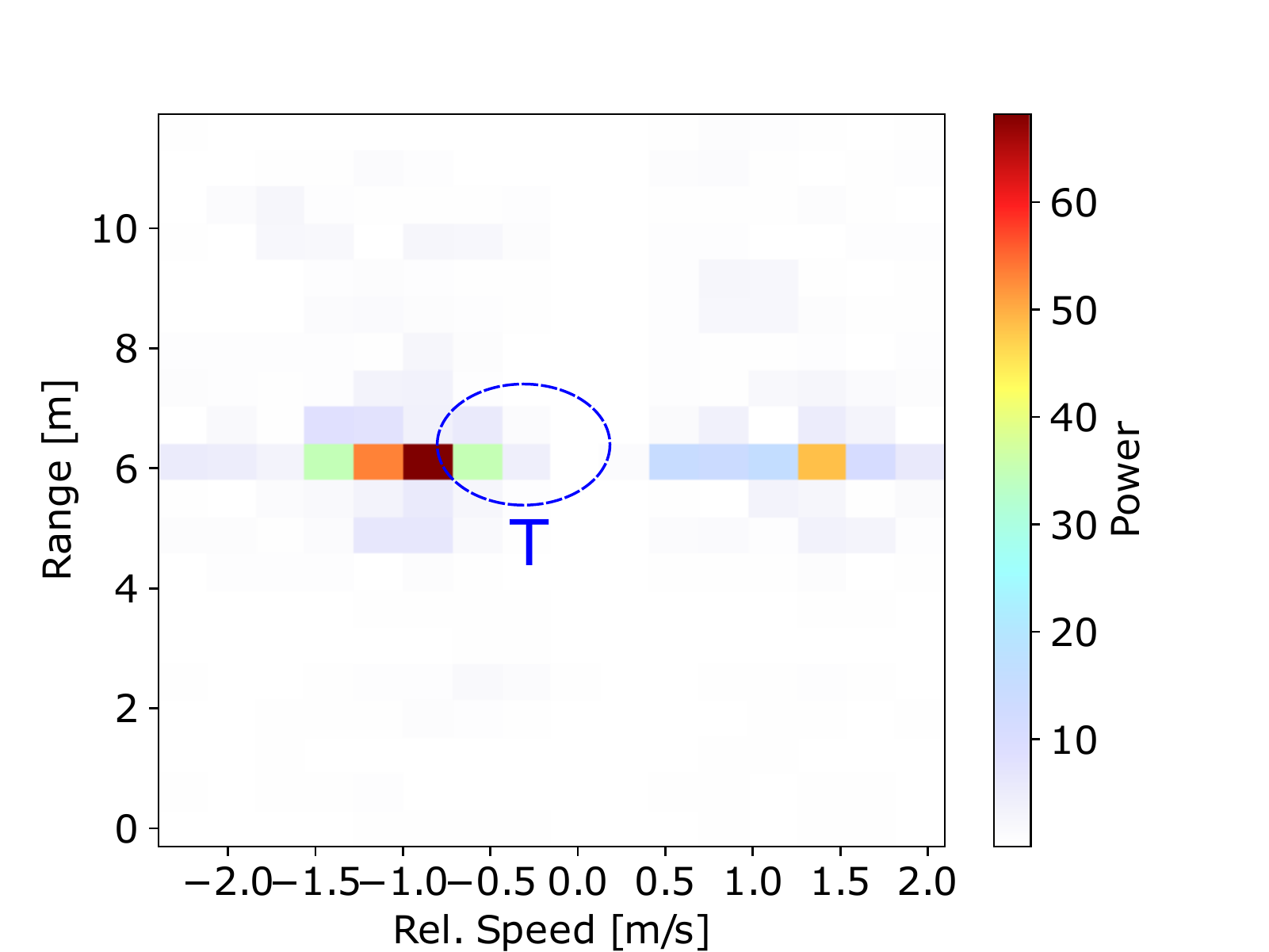} 
    \end{center}
   \caption{clutter removal with \gls{eca-c}, pedestrian close to clutter at range of ca. 6~m.}
   \label{fig:eca_c_close}
\end{subfigure}
\:
\begin{subfigure}{0.325\linewidth}
    \begin{center}
        \includegraphics        [width=1.0\linewidth,
        height = 5cm,
        clip,
        trim={5mm 0.5mm 5mm 14mm}]
        {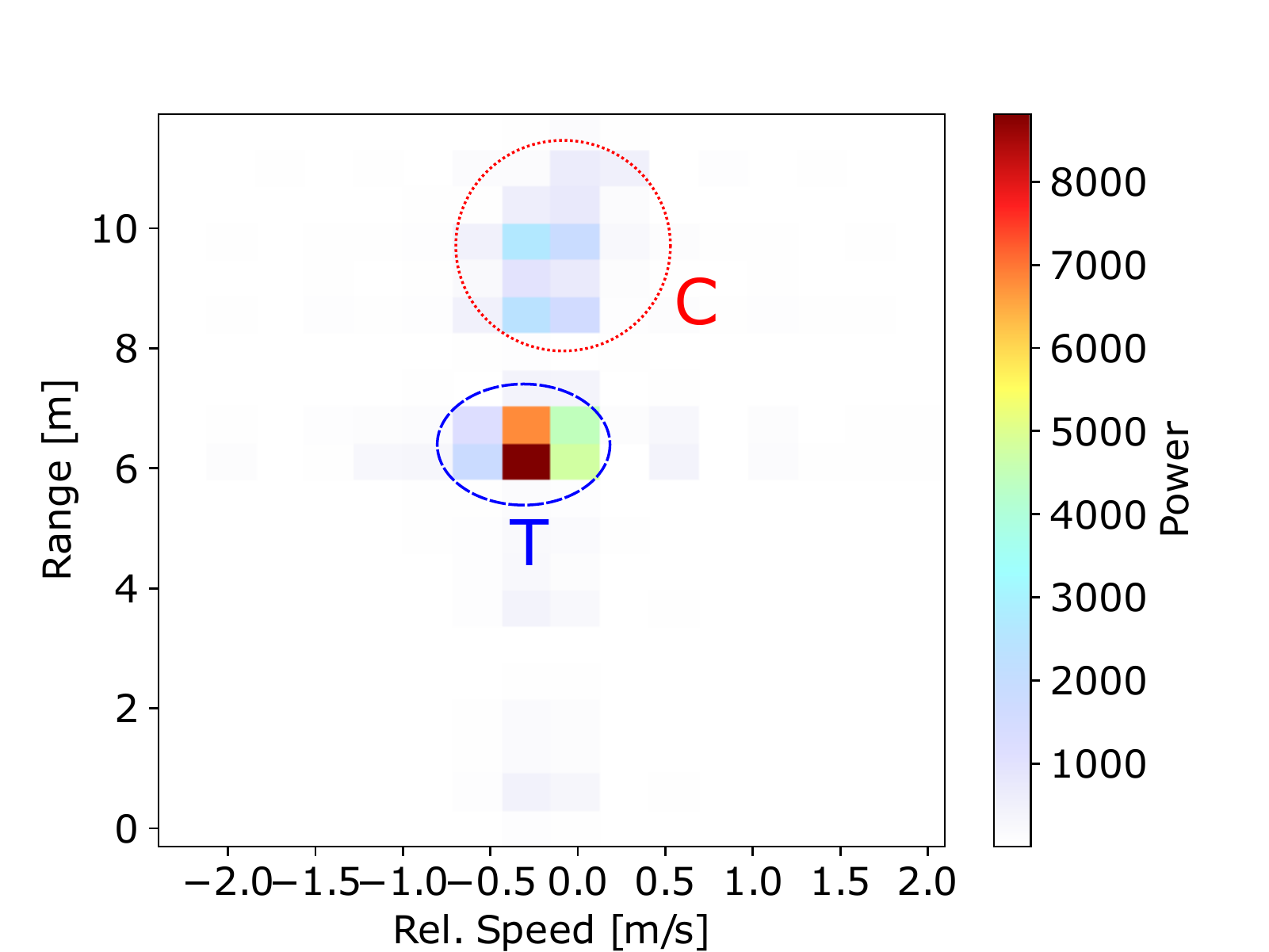} 
   \end{center}
   \caption{clutter removal with \gls{crap}, pedestrian close to clutter at range of ca. 6~m.}
   \label{fig:crap_close}
\end{subfigure}

\caption{Periodograms of sensing acquisition without applying clutter removal (left), clutter removal with ``\gls{eca-c}" (middle), and clutter removal using our proposal ``\gls{crap}" (right). In the first row, the human target~\textbf{T} is walking at a range of ca. 3~m, in the second row close to the strong clutter component~\textbf{C} at a range of ca. 6~m.. Both ``\gls{eca-c}" and ``\gls{crap}" remove the clutter well, but only ``\gls{crap}" does not create suppression at zero Doppler, enabling the detection of the slow moving target.}
\label{fig:clutter_rem_lab}
\end{figure*}
The \gls{kf} is initialized as $\mathbf{x}_0 = [\hat{r}_0, \hat{v}_0]^\text{T}$ after the first detection. Then, using a constant velocity motion model, the predicted state vector and covariance at time index $i$ are computed as
\begin{align}
\hat{\mathbf{x}}_{i|i-1} &= 
\begin{bmatrix}
1 & \Delta t \\
0 & 1 \\
\end{bmatrix} \mathbf{{x}}_{i-1|i-1} \\
\hat{\mathbf{P}}_{i|i-1} &= 
\begin{bmatrix}
1 & \Delta t \\
0 & 1 \\
\end{bmatrix} 
\mathbf{P}_{i-1|i-1}
\begin{bmatrix}
1 &  0 \\
\Delta t & 1 \\
\end{bmatrix} + \mathbf{Q}
\label{eq:kf_predict} \;
\end{align}
where $\mathbf{Q}$ is the process noise covariance and $\Delta t$ the time between measurements, which in our experiments corresponds to 10~ms (duration of one radio frame).
Assuming a linear measurement model where the state is directly observed with \gls{awgn}, the current observation $\mathbf{z}_i~=~[\hat{r}_i, \hat{v}_i]^\text{T}$ is fed to the \gls{kf} in the update step to obtain the a posteriori state vector and covariance matrix as
\begin{align}
\mathbf{x}_{i|i} &= \hat{\mathbf{x}}_{i|i-1} + \mathbf{K}_i \left(\mathbf{z}_i - \hat{\mathbf{x}}_{i|i-1} \right)  \\ 
\mathbf{P}_{i|i} &= \left( \mathbf{I} - \mathbf{K}_i  \right)\hat{\mathbf{P}}_{i|i-1}
\label{eq:kf_update} \;
\end{align}
where $\mathbf{K}_i = \hat{\mathbf{P}}_{i|i-1}  ( \hat{\mathbf{P}}_{i|i-1} + \mathbf{R}_i)^{-1}$ is the Kalman gain with $\mathbf{R}_i$ denoting the measurement noise covariance~\cite{kalman1960new}. 
\\The \gls{kf} is reset 1) in case of a timeout (time since last detection $> T_\text{max} = 0.5$~s), or 2) if either its range or speed standard deviation exceeds the pre-defined upper bound of $\sigma_{r,\text{max}} = 2$~m and $\sigma_{v,\text{max}} = 2$~m/s, respectively.

The results discussed in the following are obtained from experiments in our lab, where a pedestrian was walking from the sensing system in the direction of the strong clutter component as shown in Fig.~\ref{fig:walking_lab}. This is the underlying scenario for all \gls{poc} results discussed hereinafter.



\subsection{Clutter Removal Example}

We first demonstrate the benefits of \gls{crap} by means of range-speed periodogram visualizations of a single sensing acquisition from our \gls{isac} \gls{poc} pedestrian tracking experiments. As in Sec.~\ref{sec:sim_results}, $K=100$ clutter acquisitions were recorded. Here, however, we estimated the clutter order using \gls{mdl} for which $L=3$ was obtained. 

Fig.~\ref{fig:clutter_rem_lab} visualizes the periodograms of a single radio frame for two different scenarios: in the figures of the first row, the human target~\textbf{T} is moving at a range of ca. 3~m from the \gls{ru}, whereas the second row shows periodograms with the pedestrian walking close to the strong clutter~\textbf{C} (metal cabinet), comparable to \cref{fig:walking_lab}. One can discern that without clutter removal (Figs.~\ref{fig:no_rem_3m} and \ref{fig:no_rem_close}) the most energy is backscattered by the metal cabinet in both scenarios, which further underlines the necessity of proper clutter removal techniques in practice.
\\The periodograms after applying clutter removal with ``\gls{eca-c}" are displayed in the middle (Figs.~\ref{fig:eca_c_3m} and \ref{fig:eca_c_close}). Due to the strong static clutter, a deep null at zero Doppler is created, impacting also the slow moving pedestrian and thereby impeding its detection. While some residual target contributions remain, their energy is much lower (note the different color scales for each periodogram), rendering the detection of the pedestrian difficult. Moreover, the remaining peaks are shifted due to the attenuation at zero Doppler, which leads to an increased velocity estimation error (cf. also \cref{fig:speed_rmse}).
\\The figures in the right column (Figs.~\ref{fig:crap_3m} and \ref{fig:crap_close}) depict the periodograms after applying clutter removal according to our proposal~``\gls{crap}". The reflection caused by the pedestrian is now the strongest contribution that remains in the periodogram in both figures. \cref{fig:crap_close} demonstrates that \gls{crap} is still able to isolate a target even if it is walking slowly and very close to the strong clutter component, although compared to \cref{fig:crap_3m} more residual clutter contributions remain. This allows  the detection of the pedestrian with the peak search described in Section~\ref{sec:sim_results} without the need for processing other peaks caused by clutter.


\subsection{Pedestrian Tracking Example}

\begin{figure}[t]
\centering
  \includegraphics
  [width=1.0\linewidth,
  clip,
  trim={52mm 30mm 10mm 22mm}]
  {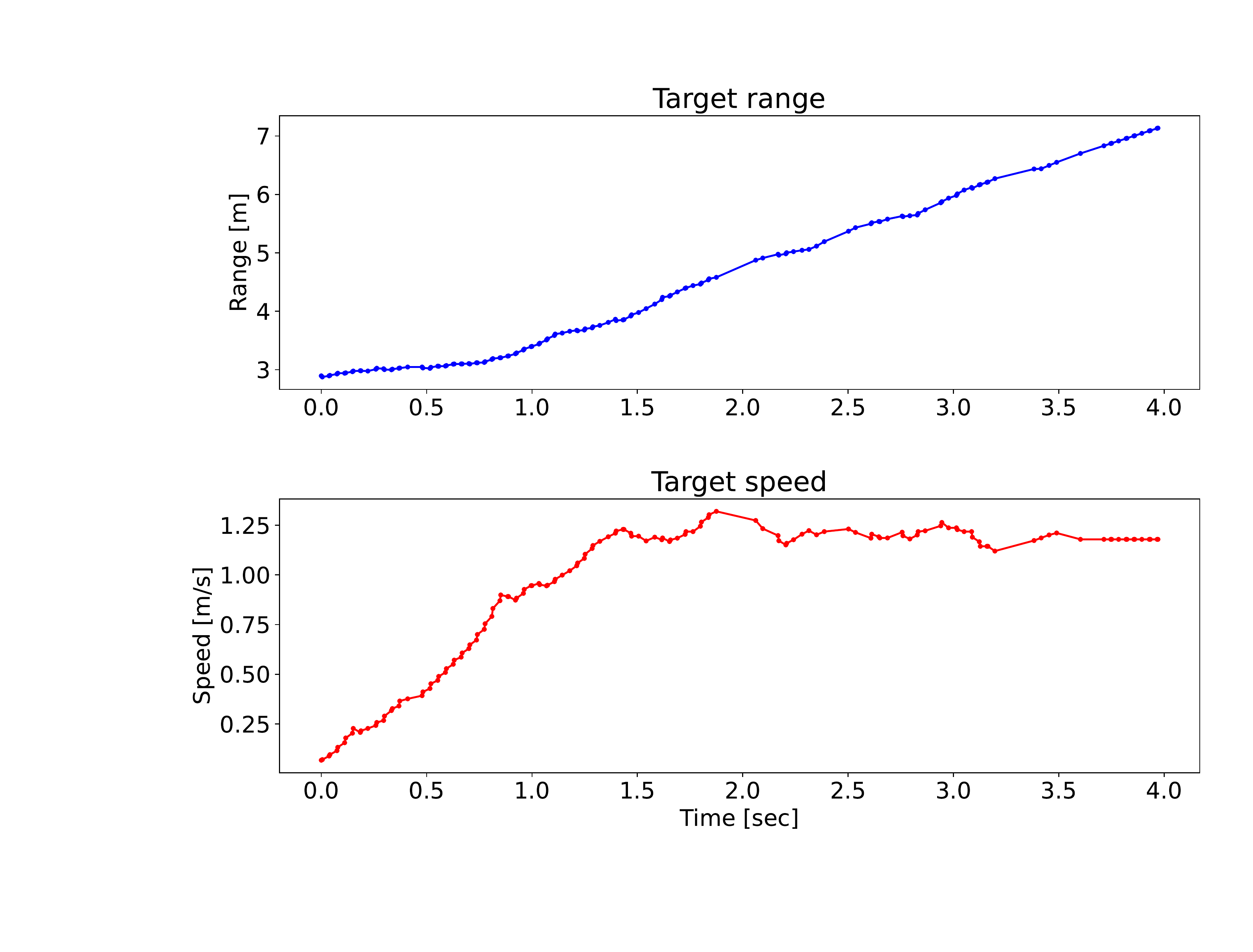}
\caption{Range (above) and speed (below) of a pedestrian tracked by the \gls{isac} \gls{poc} in a real-world lab environment. \gls{crap} is used for clutter removal and a \gls{kf} is used for post-processing the strongest peak extracted from the periodogram.}
\label{fig:kf_tracking}
\end{figure}

Finally, Fig. \ref{fig:kf_tracking} shows an excerpt with a duration of 4~s of a pedestrian's range and speed being tracked in our lab with the \gls{isac} \gls{poc}. One can observe that our system is able to track the pedestrian's movement stably without being impacted by the reflections caused by the strong clutter (metal cabinet in Fig. \ref{fig:walking_lab}) or other backscattering components thanks to the use of \gls{crap}.
\todo[inline]{How about showing also the ECA-C only tracking output? Too much effort?
\\MH: I think so, since ECA-C is not integrated in the real-time processing.
\\SM: let's be prepared for that. If I were a reviewer, this is the first thing I would ask}

\section{Conclusion}\label{sec:conclusion}

In this work, we proposed a clutter removal algorithm named \gls{crap}, which is robust to phase noise and preserves zero Doppler information. Our simulation study showed that
the missed detection probability can be reduced by at least one order of magnitude in high \gls{snr} regimes, which will likely be available in real scenarios. Moreover, we demonstrated pedestrian tracking results obtained with our \gls{isac} \gls{poc} that deploys \gls{crap} in conjunction with a \gls{kf}, which further emphasize \gls{crap}'s benefits and real-time capability. Future work should investigate decimation and interpolation techniques to further reduce the computational complexity during clutter acquisition, as well as the extension of \gls{crap} to also include the angular domain. 






\section*{Acknowledgments}
We would like to thank Frank Schaich  and {Junqing} Guan for their insights and feedback during this work's development and implementation in the \gls{isac} PoC.

This work was developed within the KOMSENS-6G project, partly funded by the German Ministry of Education and Research under grant 16KISK112K.

\bibliographystyle{IEEEtran}
\bibliography{Crap}

\end{document}